\documentclass[a4paper,twoside]{article}
\newcommand{\affil}[1]{$^{\rm #1}$}
\usepackage[authoryear]{natbib}
\bibpunct{(}{)}{;}{a}{}{,}
\usepackage{graphicx}
\date{} 
\newcommand{\kms}{\mbox{km\,s$^{-1}$}}
\newcommand{\arcsec}{$^{\prime\prime}$}
\newcommand{\arcmin}{$^\prime$}
\newcommand{\ammonia}{NH$_3$}
\newcommand{\water}{H$_2$O}
\title{\large\bf\flushleft A Pilot Survey for the H$_2$O Southern Galactic Plane Survey (HOPS)}
\author{\parbox{\textwidth}{\flushleft
\vspace{-0.5cm}
{\it Andrew J. Walsh\affil{A,B}, Nadia Lo \affil{C,D}, Michael G. Burton\affil{C}, Graeme L. White\affil{A}, Cormac R. Purcell \affil{E}, Steven N. Longmore\affil{F}, Chris J. Phillips\affil{D}, and Kate J. Brooks\affil{D}}\\
\vspace{0.4cm}
{\small \affil{A}\,Centre for Astronomy, James Cook University, Townsville, QLD, 4811, Australia}\\
{\small \affil{C}\,School of Physics, University of NSW, Sydney, NSW, 2052, Australia}\\
{\small \affil{D}\,Australia Telescope National Facility, CSIRO, PO Box 76, Epping, NSW, 1710, Australia}\\
{\small \affil{E}\,Alan Turing Building, University of Manchester, Oxford Road, Manchester, M13 9PL, UK}\\
{\small \affil{F}\,Harvard-Smithsonian Center for Astrophysics, 60, Garden Street, Cambridge, MA, 02138, USA}\\
{\small \affil{B}\,Email: andrew.walsh@jcu.edu.au}}}
\begin{document}
\twocolumn[
\begin{changemargin}{.8cm}{.5cm}
\begin{minipage}{.9\textwidth}
\vspace{-1cm}
\maketitle
%
%
\small{\bf Abstract:}
We describe observations with the Mopra radiotelescope designed to assess
the feasibility of the H$_2$O maser southern Galactic plane survey (HOPS).
We mapped two one-square-degree regions along the Galactic plane using
the new 12\,mm receiver and the UNSW Mopra spectrometer (MOPS).
We covered the entire spectrum between 19.5 and 27.5\,GHz using this setup with
the main aims of finding out which spectral lines can be detected with a quick
mapping survey. We report on detected emission from H$_2$O masers, NH$_3$
inversion transitions (1,1), (2,2) and (3,3), HC$_3$N (3-2), as well as several
radio recombination lines. 


\medskip
\medskip
\end{minipage}
\end{changemargin}
]
\small

\section{Introduction}
The Mopra radiotelescope is a 22-m single dish situated near Siding Spring
mountain in New South Wales, Australia. Recent upgrades of the telescope
included a new 12\,mm receiver which operates in the frequency range of 16
to 27.5\,GHz. In addition to this, the UNSW Mopra spectrometer
(MOPS) allows
instantaneous bandwidth coverage of over 8\,GHz (most of the 12\,mm band)
sampled by 32\,768 channels. Thus, MOPS potentially allows many spectral lines
to be observed simultaneously. These new hardware, together with the
on-the-fly mapping capability of the telescope,
allow for very efficient multiple-line surveys of the southern sky.

Previous observations at 3\,mm \citep{bains06} have shown that the system
works well over small regions, but it is the purpose of this work to test the
system when mapping degree-scale regions of the Galaxy at 12\,mm. Our
ultimate aim is to map a large portion of the southern Galaxy in a survey we
call ``HOPS'' (the H$_2$O Plane Survey). The brightest
spectral lines in this part of the spectrum are the H$_2$O (6$_{16}$--5$_{23}$)
maser line at 22.235\,GHz and NH$_3$ inversion transitions, which
form the main focus of our survey. The 22.235\,GHz \water~masers are
commonly found towards regions of both low and high mass star formation
\citep{genzel79}. They have also been found in 
planetary nebulae \citep{miranda01}, Mira variables
\citep{hinkle79}, Asymptotic Giant Branch (AGB) stars \citep{barlow96},
and the centres of active galaxies \citep{claussen84,braatz96},
but tend to be much
fainter than in star forming regions. Thus, HOPS will provide an untargetted
survey of \water~masers in star forming regions over a substantial portion
of our Galaxy. Comparison of the results of HOPS with the untargetted
methanol multibeam survey \citep{cohen07} will be useful in identifying the
relative prevalence of the two maser species, as well as which stages
of high mass star formation they are associated with. HOPS will also allow
us to compare the relative occurrence of \water~masers in regions of high
and low mass star formation.

\ammonia~is a
tracer of dense quiescent gas. The (1,1) inversion transition typically
exhibits prominent hyperfine structure, which can be used to infer the optical
depth of the transition. Comparison of the (1,1) and
higher J inversion transitions can be used to estimate the rotational
temperature of the gas. A survey
of \ammonia~will trace the dense gas structure of our Galaxy, since it has
an effective critical density of $\sim 10^5$\,cm$^{-3}$. \ammonia~is also known
to occur in cold regions of gas, where more common gas tracers, like CO
tend to deplete by freeze out onto dust grains \citep{bergin06}.
This makes \ammonia~a very useful tracer of the cold,
dense regions of our Galaxy. The results of HOPS can be used to identify
regions where the earliest stages of high mass star formation are taking place,
as well as their broad physical properties such as average temperature,
density and mass.

\section{Observations}
We used the Mopra Radiotelescope to map two one-square-degree regions located
at Galactic longitudes of 305.5$^\circ$ and 333.2$^\circ$. We centred each map slightly
off the Galactic plane in order to assess how far off the Galactic plane we
might find emission. Thus, the centres of the maps were (305.5$^\circ$, +0.5$^\circ$)
and (333.2$^\circ$, -0.7$^\circ$). Adjacent scans were separated by 
51\arcsec~, to give nyquist sampling of the beam FWHM (2\arcmin) at the
highest observing frequency (27.5\,GHz). At lower frequencies, the observations
consequently oversample the mapped region.
The scanning rate was 15\,arcsec\,s$^{-1}$ and spectra were stored
every 2 seconds, giving a 30\arcsec~spacing between spectra in each row.
The 2 second dump rate is the fastest that the telescope software can used,
which limits the speed at which any map can be made. We could increase this 
speed by a factor of two so that the the data is dumped every half-beam, but
prefer the slower speed to increase the signal to noise ratio.
We mapped each region twice, scanning once in Galactic
longitude and once in Galactic latitude and averaged each data
cube. This method helps to reduce noise
levels, as well as minimise artificial stripes that are introduced when only
one scanning direction is used. It also allows us to carefully check spurious
emission that may occur in one scan, but not the other. Each scan was
completed in approximately 8 hours.

Observations were conducted in the first week of December 2006, during the
Australian summer. We chose this time of the year partly to assess the effects
of observing during summer. During the observations, the telescope was
subjected to conditions ranging from hot, dry days, with temperatures rising
as high as 310\,K (37$^\circ$C) to cool evenings, from clear
skies to thin cloud cover to storms. We continued observations in
all conditions except for storms to see how the data were affected.

The broadband mode of MOPS was used, with the frequency range covering between
19.5 and 27.5\,GHz. Table \ref{tab1} lists the brighter line transitions
within this range. The channel width varied between 4.1\kms~at 19.5\,GHz and
2.9\kms~at 27.5\,GHz. This is usually insufficient to resolve Galactic
thermal and maser lines as line widths are typically a few \kms~or less.

\begin{table}
\caption{\protect\footnotesize{Lines in the 19.5-27.5\,GHz range.
Lines marked in bold are the main target lines for HOPS.}}
\scriptsize{
\begin{tabular}{lccc}
\hline
Line & Freq. & Pilot Survey & Maser or\\
Name & (GHz)     & Detection& Thermal?$^a$\\
\hline
H69$\alpha$                            & 19.591    &  Y   & thermal\\
CH$_3$OH ($2_{1 } - 3_{0 }$)           & 19.967    &  N   & both (II)\\
H68$\alpha$                            & 20.462    &  Y   & thermal\\
NH$_3$ (8,6)                           & 20.719    &  N   & both\\
NH$_3$ (9,7)                           & 20.735    &  N   & thermal\\
C$_6$H ($15/2 - 13/2$)                 & 20.792    &  N   & thermal\\
NH$_3$ (7,5)                           & 20.804    &  N   & thermal\\
NH$_3$ (11,9)                          & 21.071    &  N   & both\\
NH$_3$ (4,1)                           & 21.134    &  N   & thermal\\
H67$\alpha$                            & 21.385    &  Y   & thermal\\
{\bf H$_2$O ($6_{1 } - 5_{2 }$)}       &{\bf 22.235}&{\bf Y}&{\bf maser}\\
CCS ($2_1 - 1_0$)                      & 22.344    &  N   & thermal\\
H66$\alpha$                            & 22.364    &  Y   & thermal\\
HC$_7$N (20--19)                       & 22.559    &  N   & thermal\\
NH$_3$ (2,1)                           & 23.099    &  N   & thermal\\
CH$_3$OH ($9_{2  } - 10_{1   }$)       & 23.121    &  N   & maser (II)\\
H65$\alpha$                            & 23.404    &  Y   & thermal\\
CH$_3$OH ($10_{1  } - 9_{2  }$)        & 23.444    &  N   & thermal\\
{\bf NH$_3$ (1,1)}                     &{\bf 23.694}&{\bf Y}&{\bf thermal}\\
{\bf NH$_3$ (2,2)}                     &{\bf 23.722}&{\bf Y}&{\bf thermal}\\
{\bf NH$_3$ (3,3)}                     &{\bf 23.870}&{\bf Y}&{\bf both}\\
NH$_3$ (4,4)                           & 24.139    &  N   & thermal\\
NH$_3$ (5,5)                           & 24.533    &  N   & both\\
NH$_3$ (6,6)                           & 25.056    &  N   & both\\
H64$\alpha$                            & 24.509    &  Y   & thermal\\
CH$_3$OH ($3_{2 } - 3_{1 }$)           & 24.928    &  N   & both (I)\\
CH$_3$OH ($4_{2 } - 4_{1 }$)           & 24.933    &  N   & both (I)\\
CH$_3$OH ($2_{2 } - 2_{1 }$)           & 24.934    &  N   & both (I)\\
CH$_3$OH ($5_{2 } - 5_{1 }$)           & 24.959    &  N   & both (I)\\
CH$_3$OH ($6_{2 } - 6_{1 }$)           & 25.018    &  N   & both (I)\\
CH$_3$OH ($7_{2 } - 7_{1 }$)           & 25.124    &  N   & both (I)\\
{\bf H63$\alpha$}                      &{\bf 25.686}&{\bf Y}&{\bf thermal}\\
HC$_5$N (10--9)                        & 26.626    &  N   & thermal\\
H62$\alpha$                            & 26.939    &  Y   & thermal\\
{\bf HC$_3$N (3--2)}                   &{\bf 27.294}&{\bf Y}&{\bf thermal}\\
CH$_3$OH ($13_{2   } - 13_{1   }$)     & 27.472    &  N   & thermal\\
NH$_3$ (9,9)                           & 27.477    &  N   & thermal\\
&&&\\
\hline
\end{tabular}
\medskip\\
$^a$For CH$_3$OH masers, they are identified either as Class I or II\\
}
\label{tab1}
\end{table}

Data were reduced using the ATNF dedicated packages ``{\sc livedata}'' and
``{\sc gridzilla}'' developed by Mark
Calabretta\footnote{http://www.atnf.csiro.au/people/mcalabre/livedata.html}.
{\sc Livedata} performs a bandpass calibration for each row, using the preceding
{\sc off} scan. A 1$^{\rm st}$ order polynomial (ie. a straight line) is
the fit to the baseline. {\sc Gridzilla} regrids and combines the data from
multiple scanning directions onto a data cube with pixels 30\arcsec $\times$
30\arcsec. The data are also weighted according to the relevant T$_{\rm sys}$.

\section{Results}
\begin{figure}[h]
\begin{center}
\includegraphics[width=0.45\textwidth, angle=0]{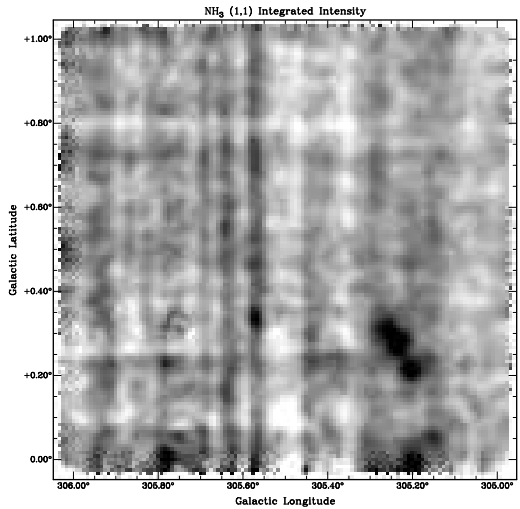}
\caption{\ammonia~(1,1) integrated intensity map for G305.5+0.5. Prominent
horizontal and vertical stripes are the result of observing during
bad weather. Real \ammonia~emission can be seen around G305.2+0.25.}
\label{fig1}
\end{center}
\end{figure}

Data cubes were made for each of the lines listed in Table \ref{tab1}.
Emission was detected in the \water, \ammonia, radio recombination lines
(H69$\alpha$ to H62$\alpha$) and
HC$_3$N data cubes. The integrated intensity map for \ammonia~(1,1) is shown
in Figure \ref{fig1} for G305.5+0.5 as an example. The prominent horizontal
and vertical stripes seen in the Figure result from observations
made in inclement weather. The
vertical stripes are most prominent between Galactic longitudes of 305.4$^\circ$ to
305.7$^\circ$ and 305.9$^\circ$ to 306$^\circ$. The horizontal stripes are most prominent
around Galactic latitudes of +0.1$^\circ$, +0.28$^\circ$ and +0.53$^\circ$.
Unfortunately, the striping makes it difficult to distinguish real
astrophysical emission in the integrated intensity map.

\begin{figure*}[h]
\begin{center}
\begin{tabular}{cc}
\includegraphics[width=0.45\textwidth, angle=0]{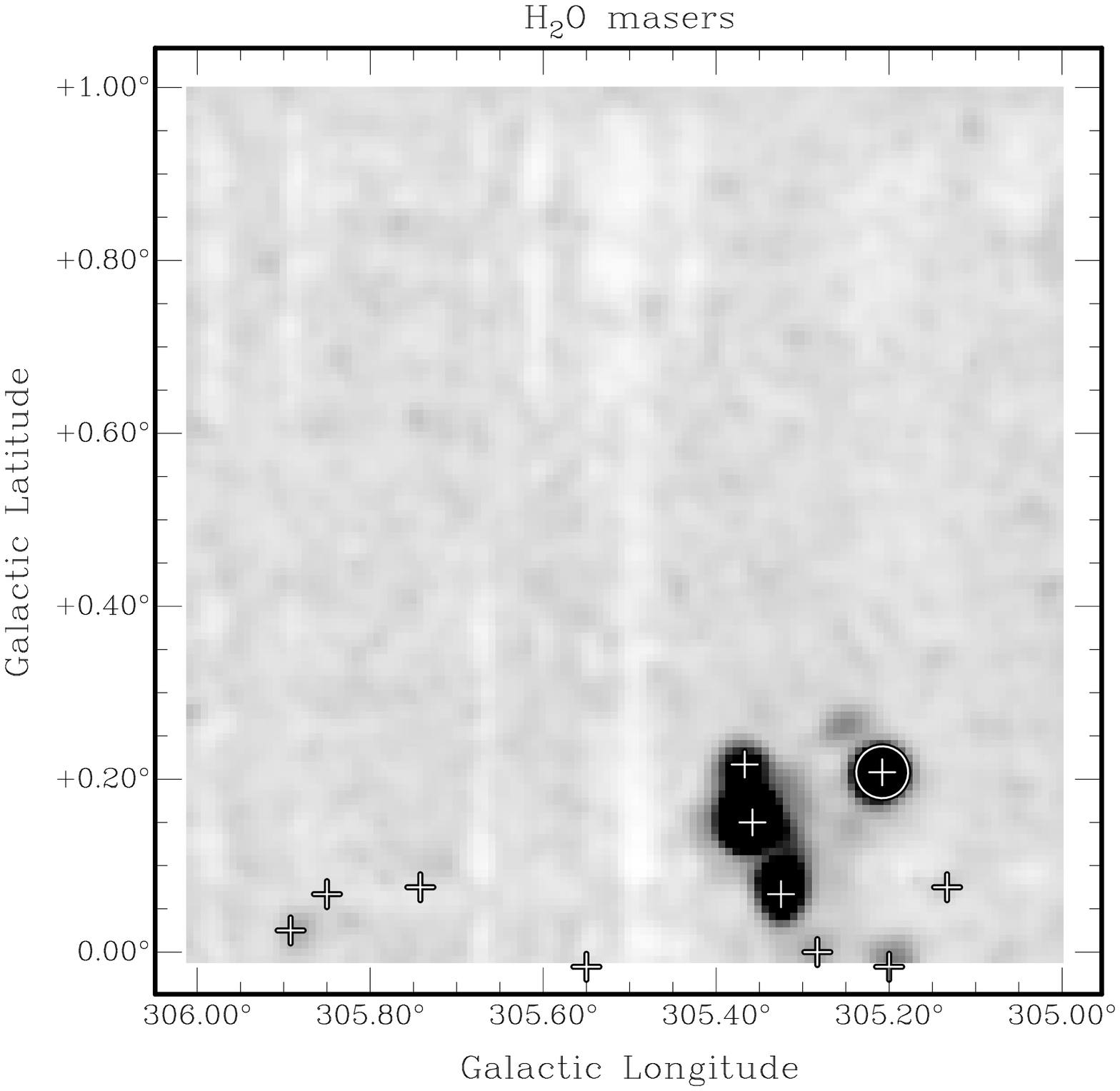}&
\includegraphics[width=0.45\textwidth, angle=0]{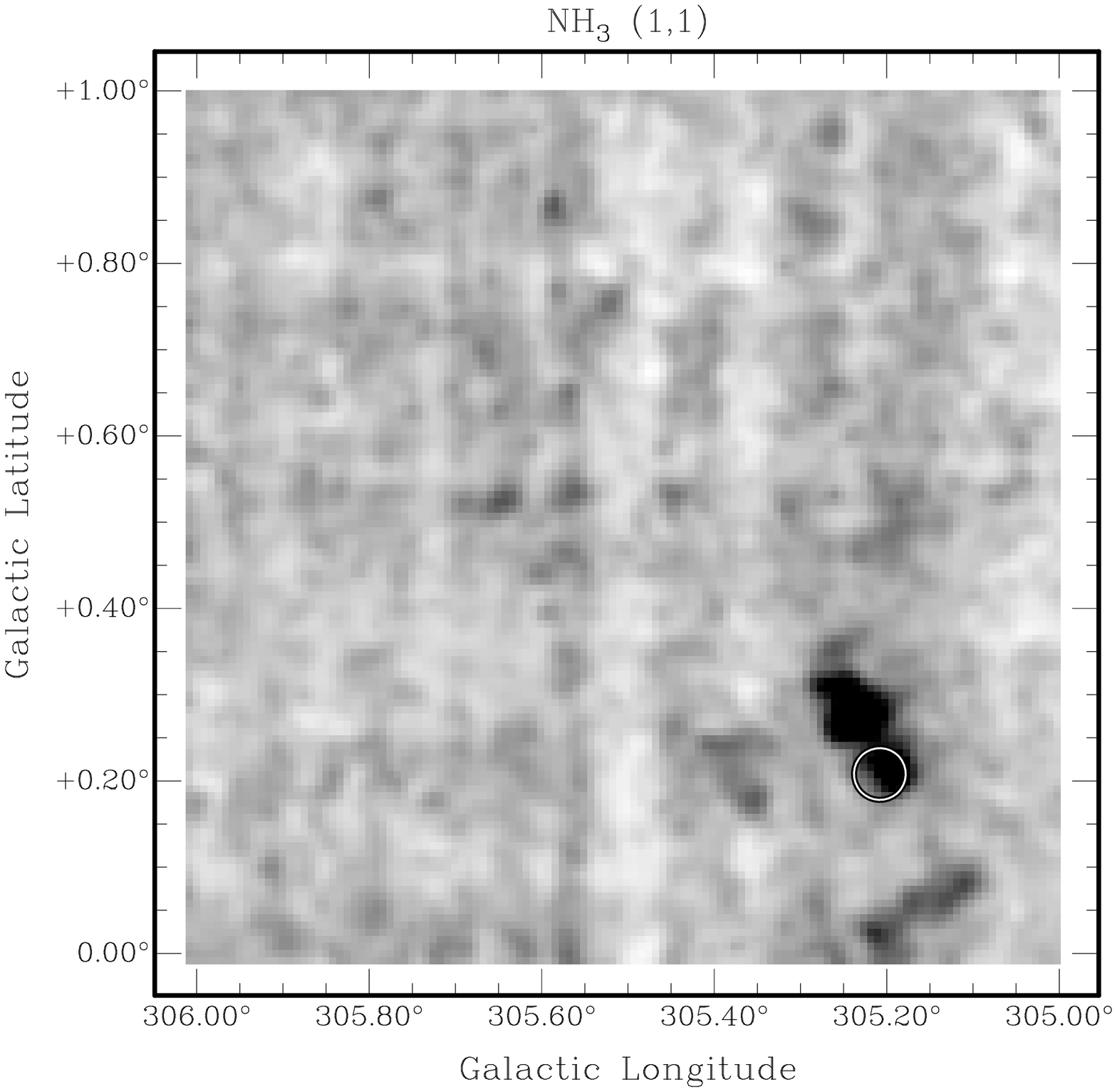}\\
\includegraphics[width=0.45\textwidth, angle=0]{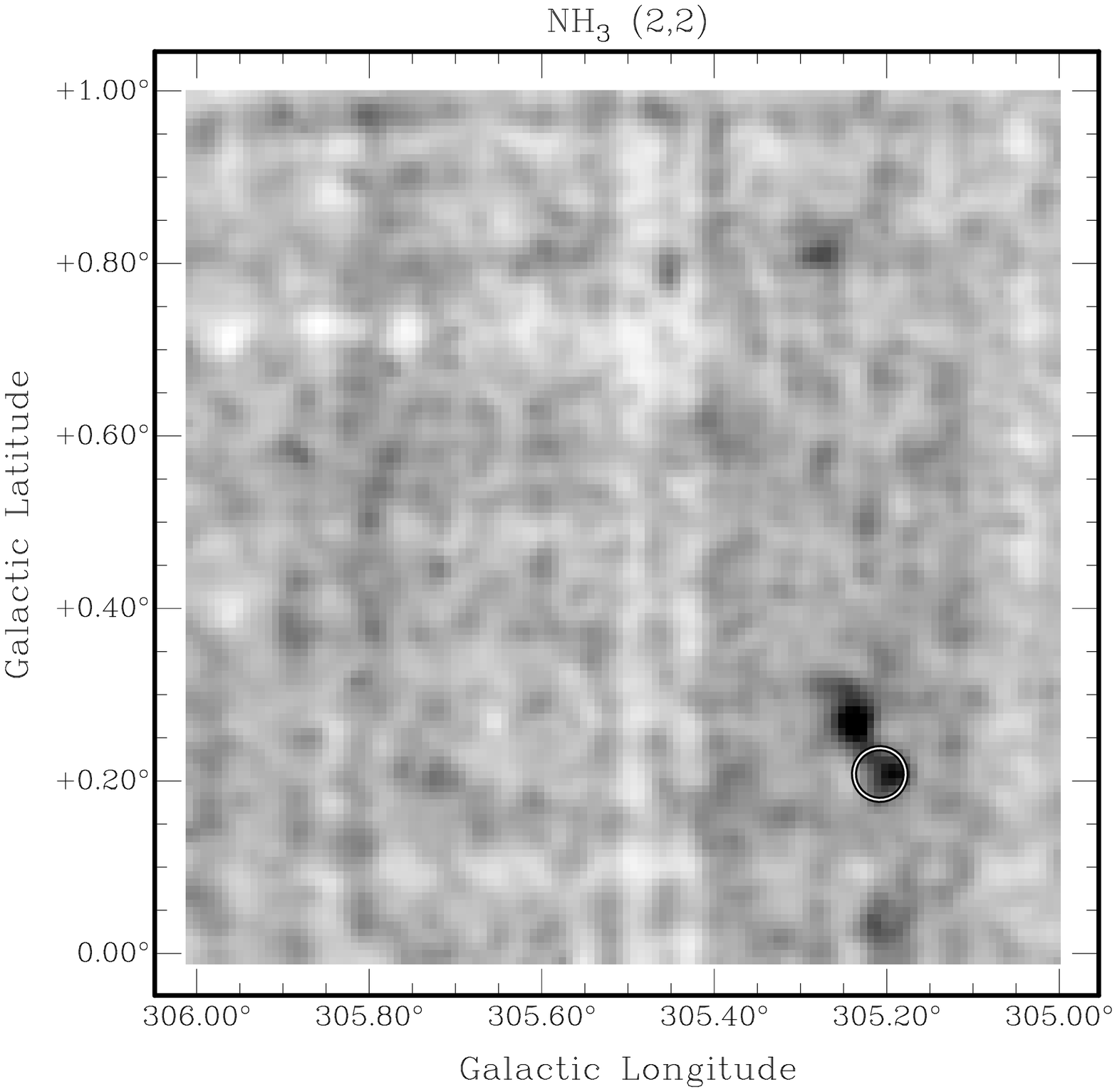}&
\includegraphics[width=0.45\textwidth, angle=0]{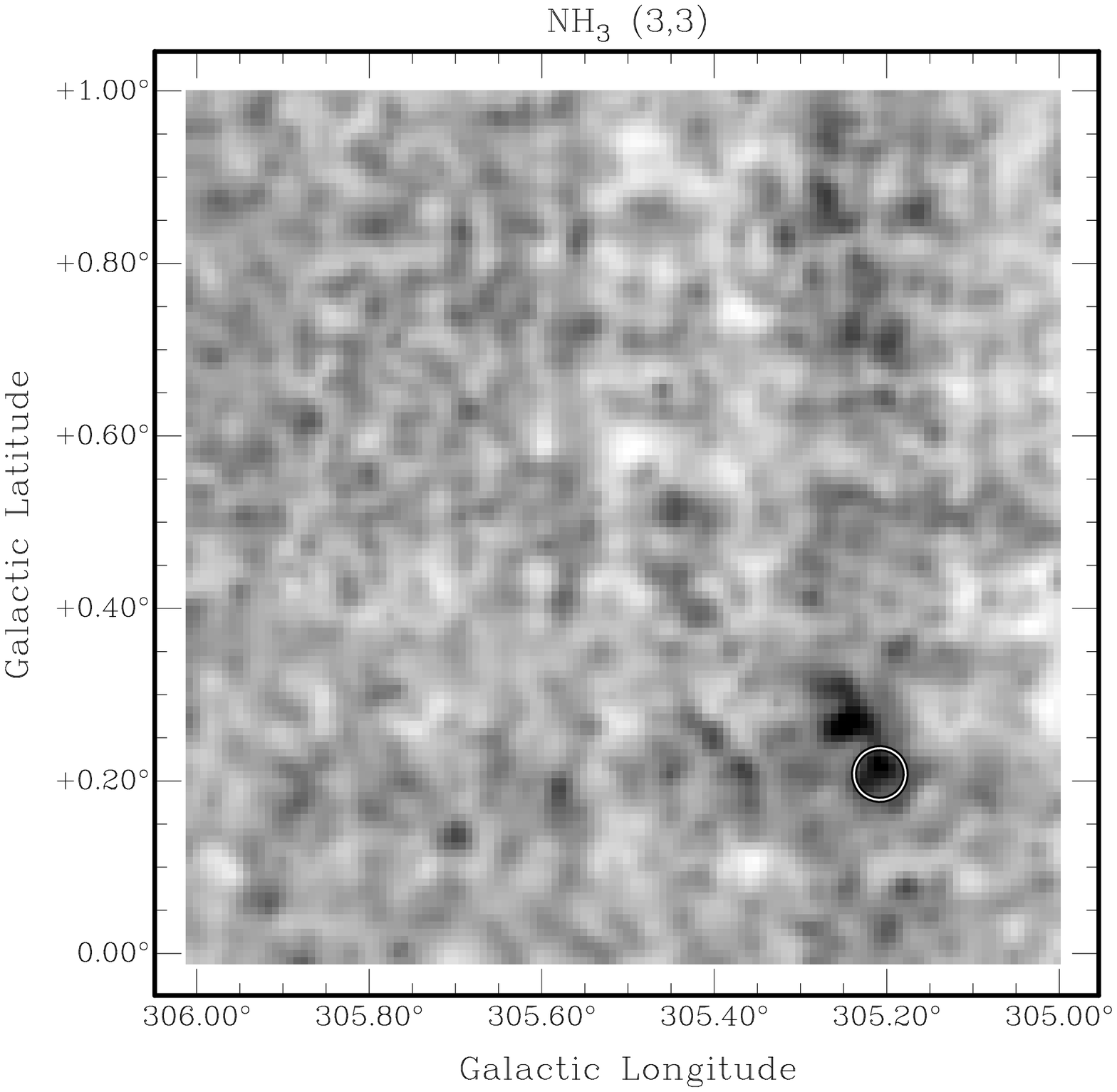}\\
\includegraphics[width=0.45\textwidth, angle=0]{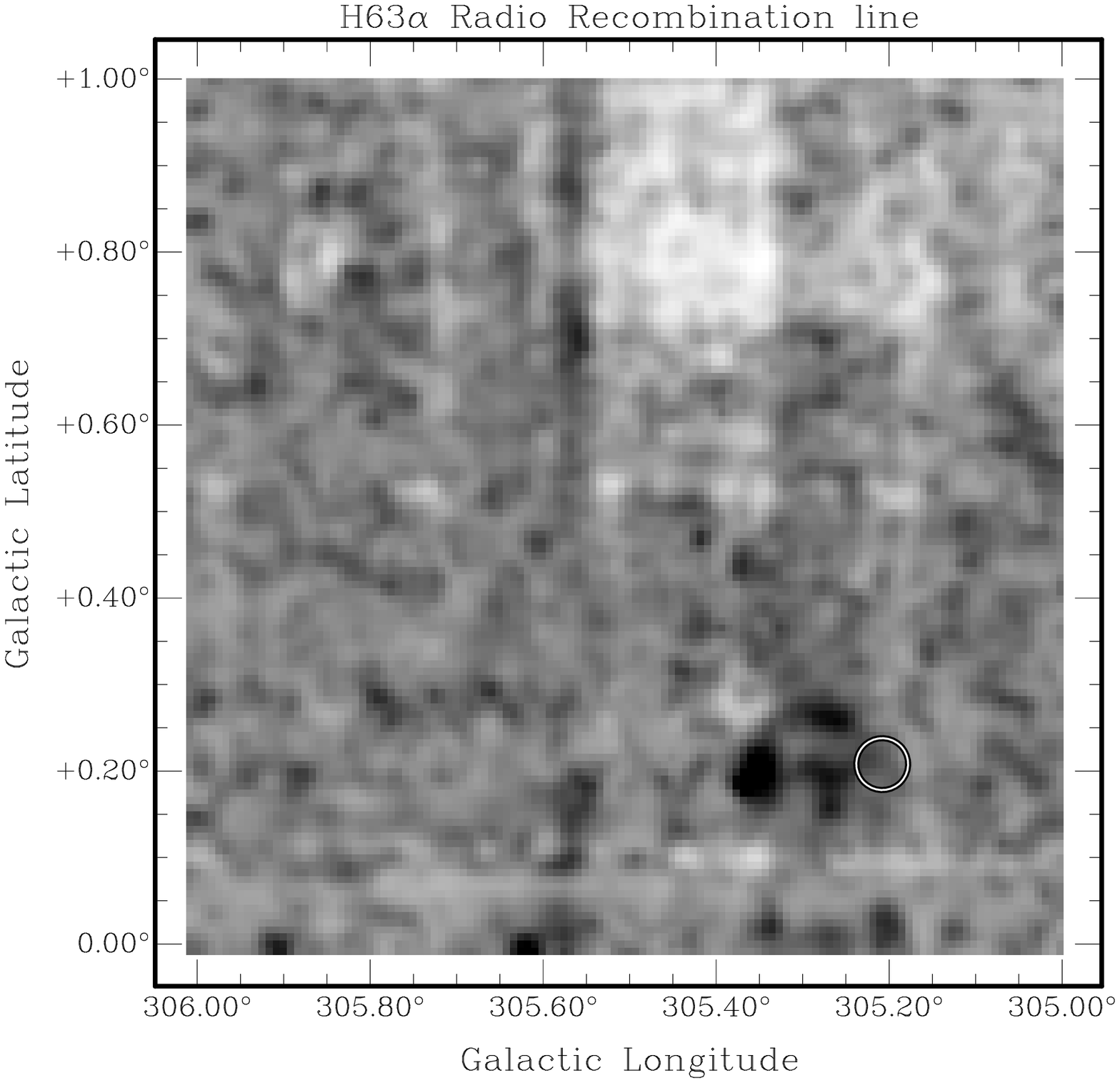}&
\includegraphics[width=0.45\textwidth, angle=0]{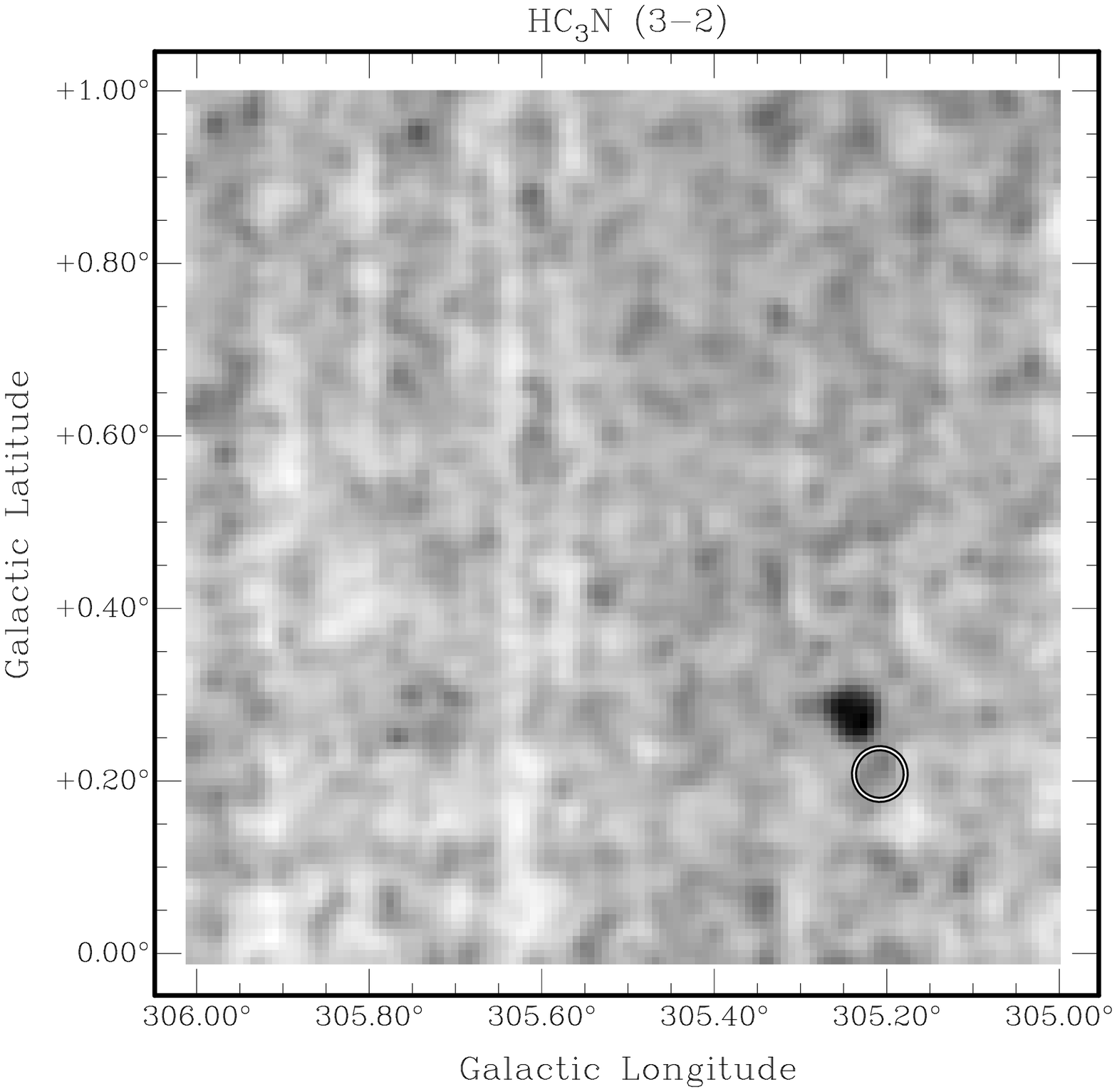}\\
\end{tabular}
\caption{Peak intensity to noise ratio images for G305.5+0.5. These images show the
ratio of the peak intensity to rms noise level derived from the spectrum
at each pixel. They show where real emission occurs and down-weight
the contribution due to noisy data. Top-left is the image
for \water~masers, top-right is \ammonia~(1,1), middle-left is
\ammonia~(2,2), middle-right is \ammonia~(3,3), bottom-left is H63$\alpha$
radio recombination line and bottom-right is HC$_3$N (3--2). The plus symbols
on the \water~maser image indicate the positions of detected \water~masers. The
circle on the \water~maser image shows the position of spectra shown in Figure
\ref{fig3}.}
\label{fig2a}
\end{center}
\end{figure*}

\begin{figure*}[h]
\begin{center}
\begin{tabular}{cc}
\includegraphics[width=0.45\textwidth, angle=0]{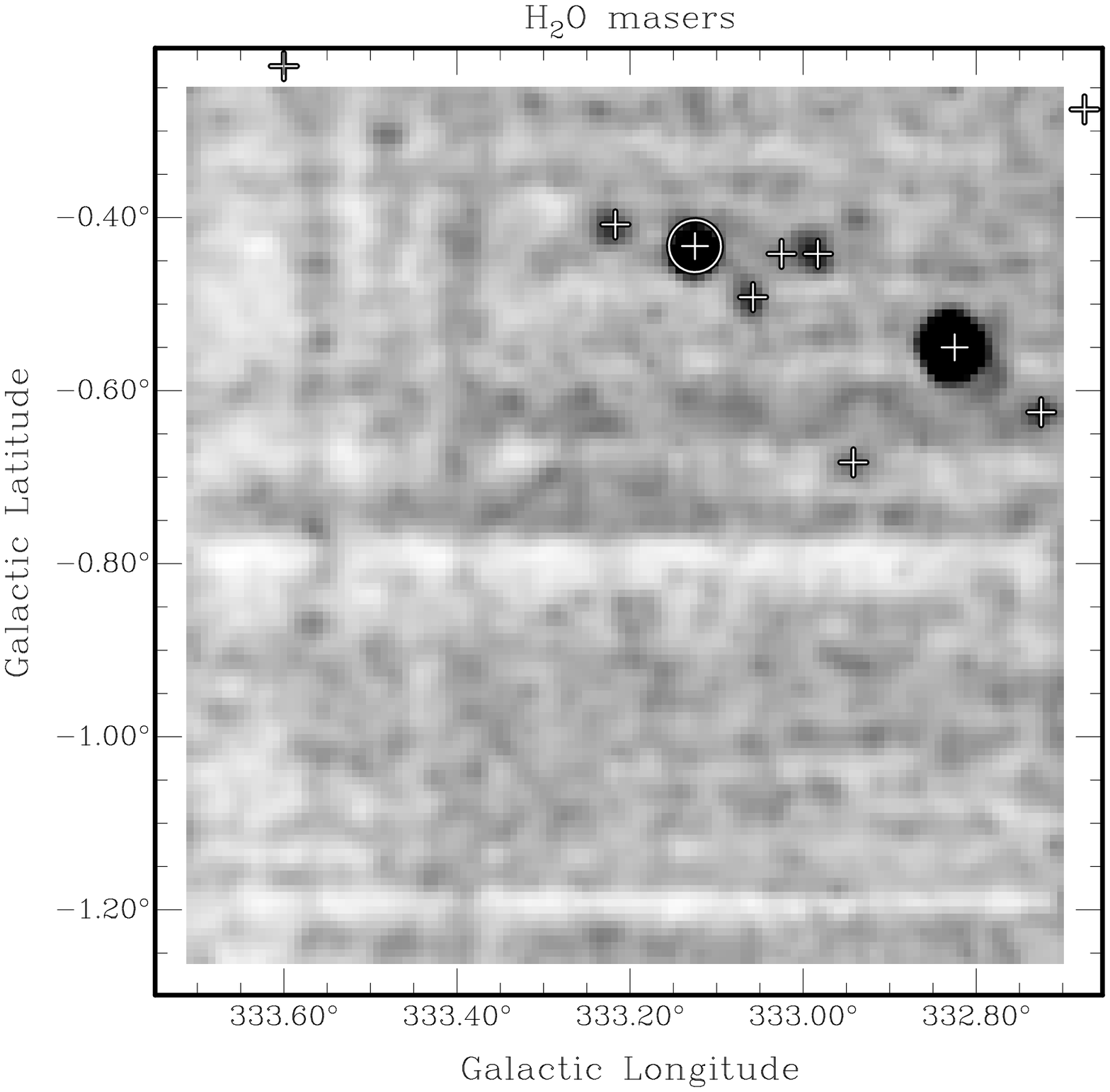}&
\includegraphics[width=0.45\textwidth, angle=0]{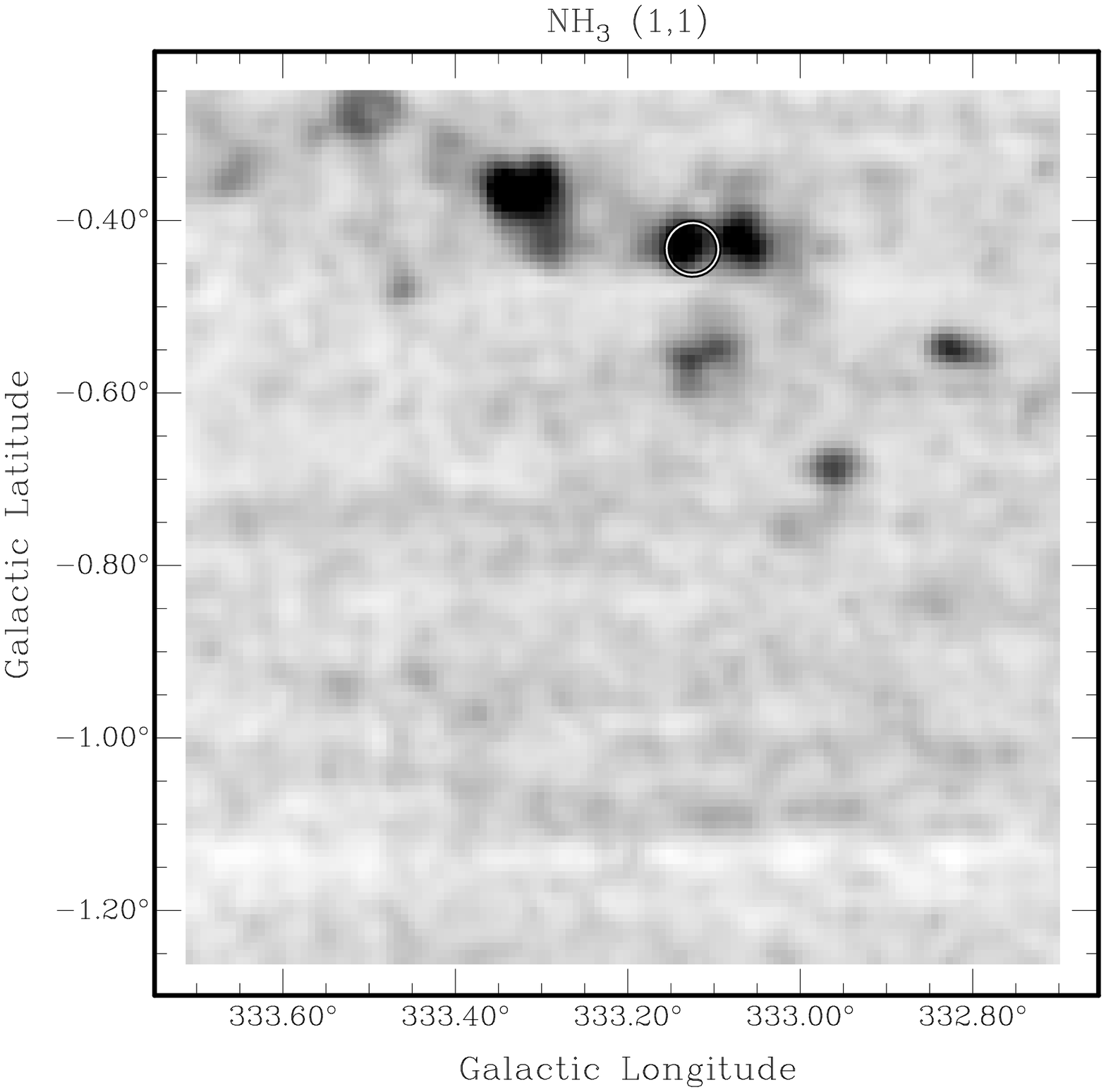}\\
\includegraphics[width=0.45\textwidth, angle=0]{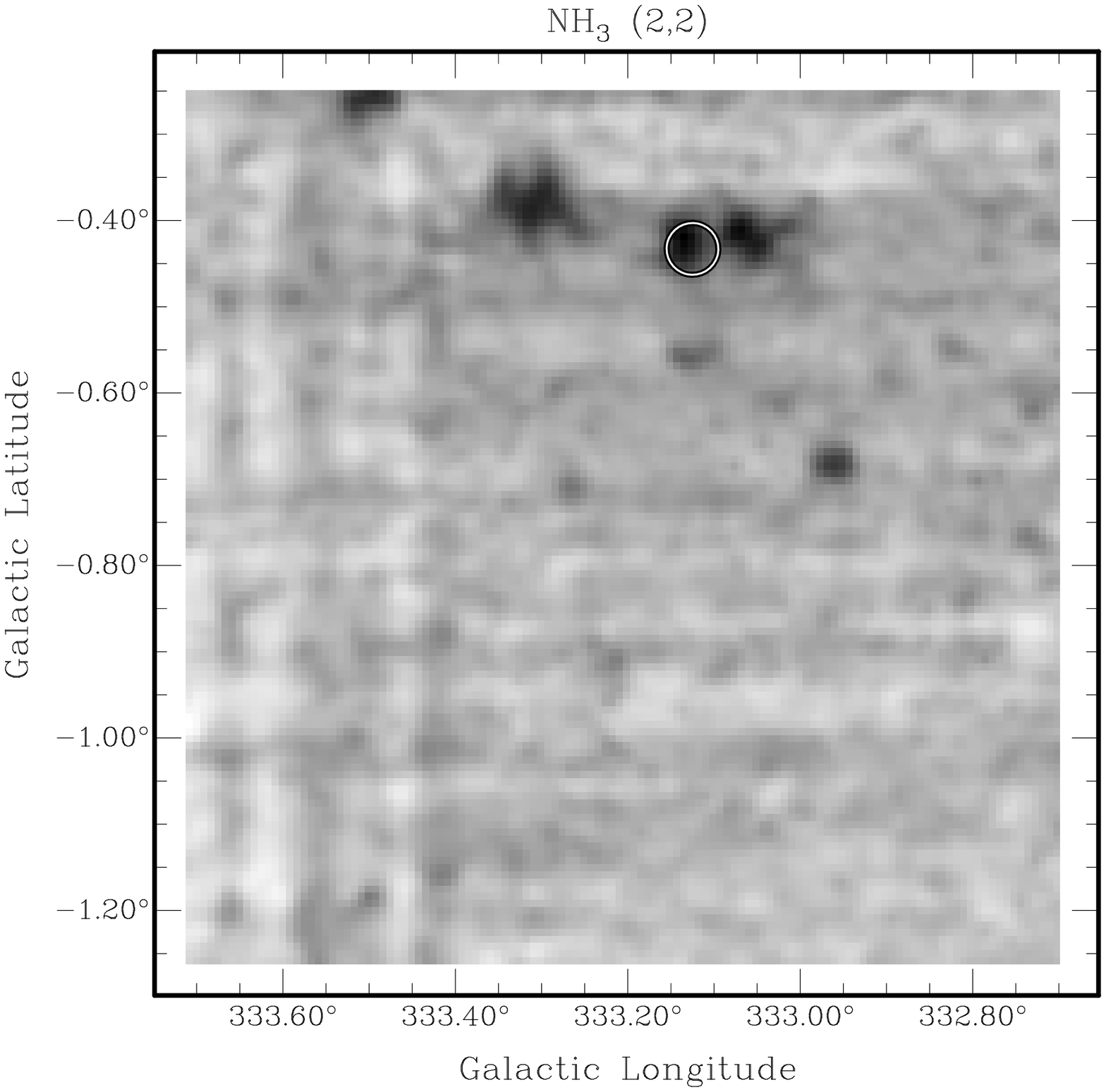}&
\includegraphics[width=0.45\textwidth, angle=0]{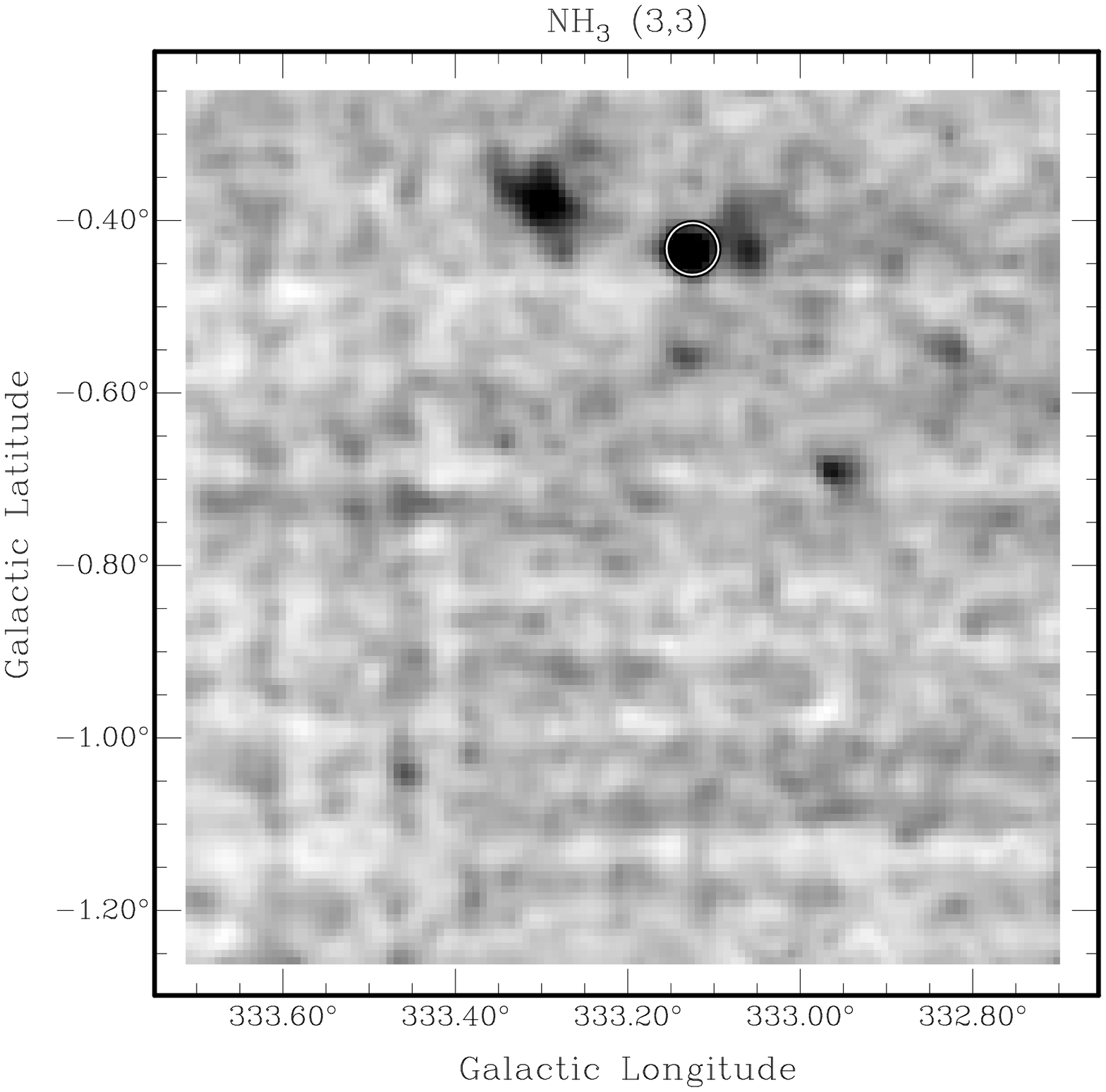}\\
\includegraphics[width=0.45\textwidth, angle=0]{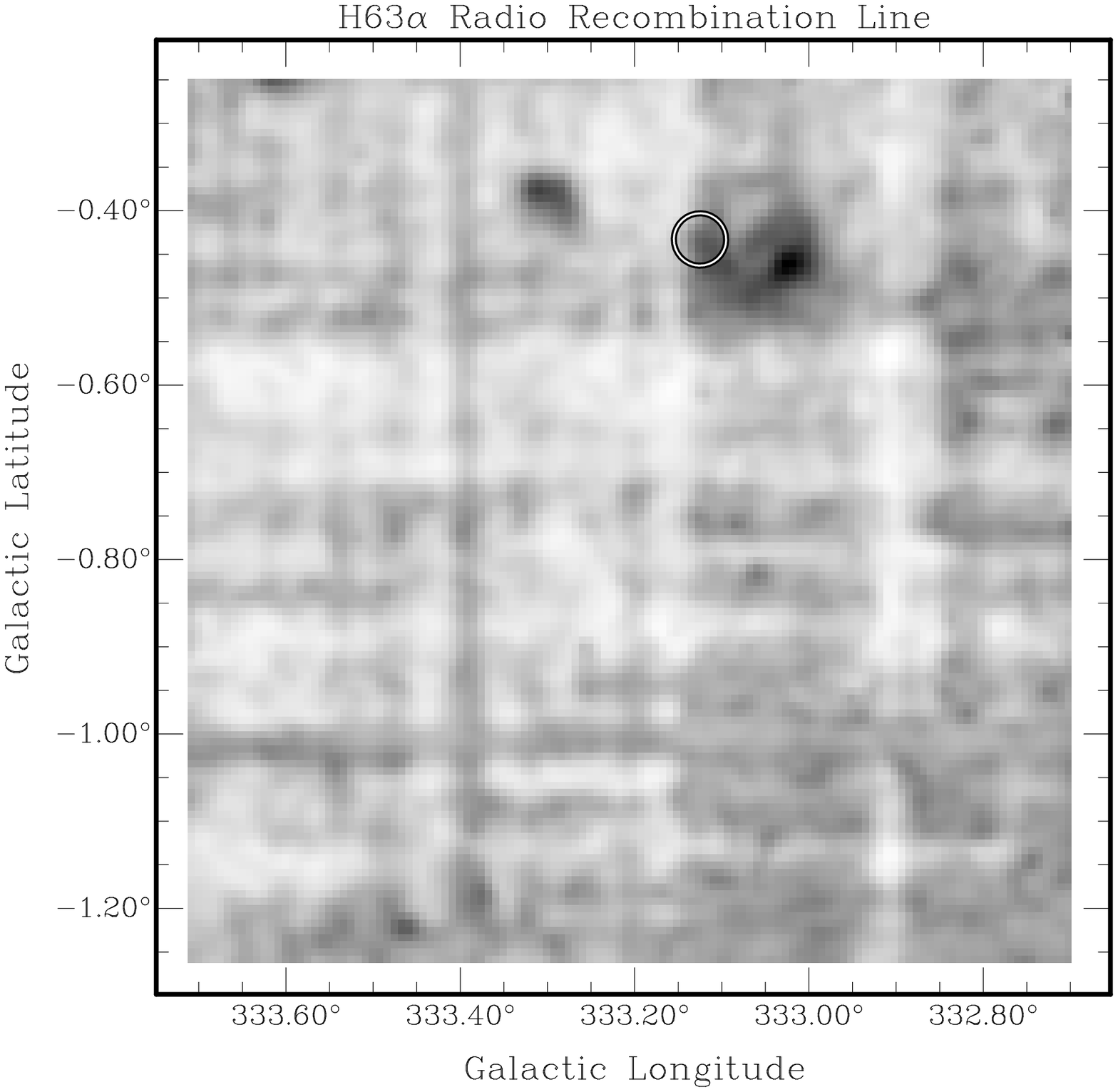}&
\includegraphics[width=0.45\textwidth, angle=0]{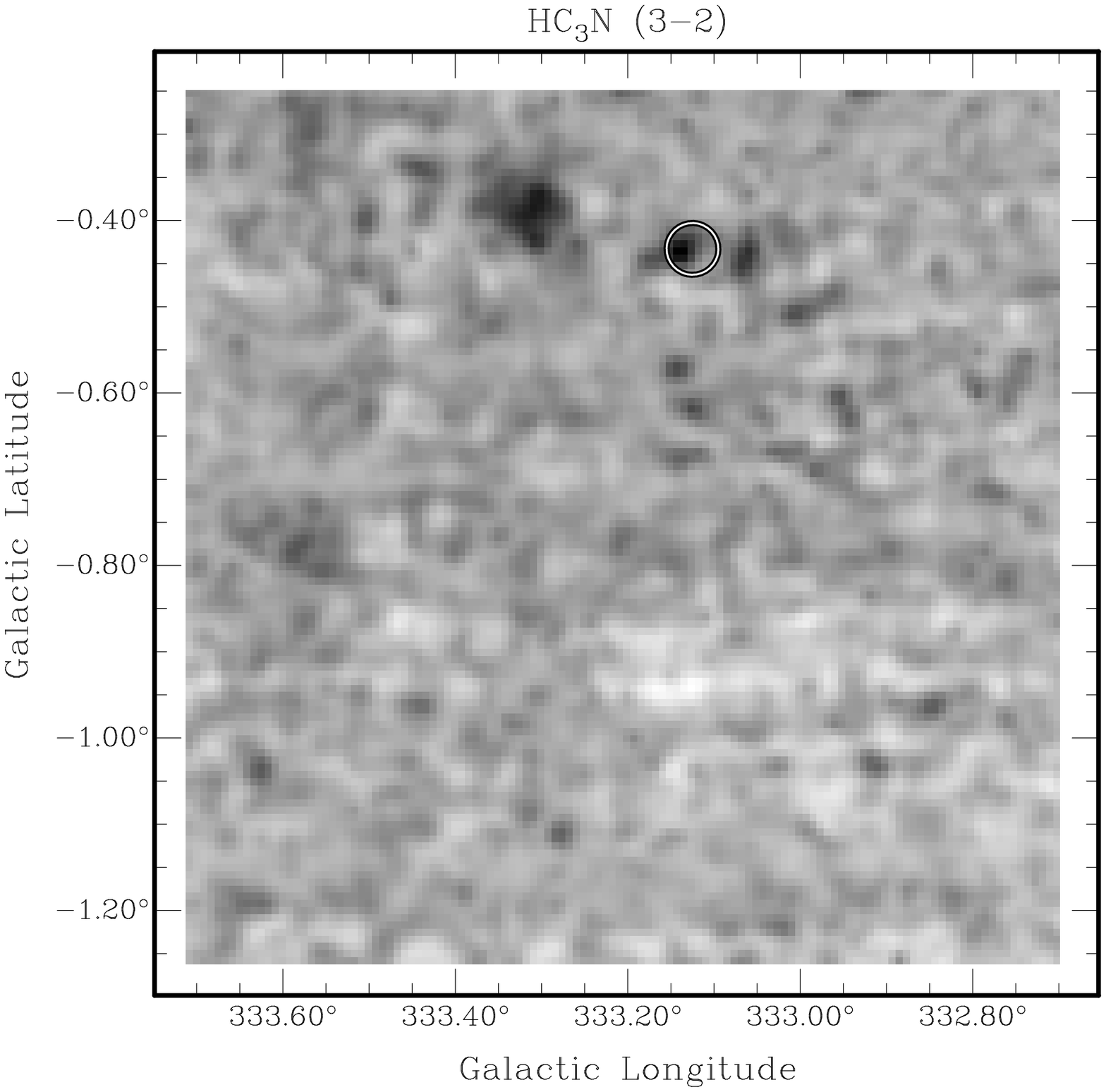}\\
\end{tabular}
\caption{Peak intensity to noise ratio images for G333.2-0.7. These images show the
ratio of the peak intensity to rms noise level derived from the spectrum
at each pixel. They show where real emission occurs and down-weight
the contribution due to noisy data. Top-left is the image
for \water~masers, top-right is \ammonia~(1,1), middle-left is
\ammonia~(2,2), middle-right is \ammonia~(3,3), bottom-left is H63$\alpha$
radio recombination line and bottom-right is HC$_3$N (3--2). The plus symbols
on the \water~maser image indicate the positions of detected \water~masers. The
circle on the \water~maser image shows the position of spectra shown in Figure
\ref{fig4}.}
\label{fig2b}
\end{center}
\end{figure*}

Bad weather introduces noise into each spectrum and affects all frequencies
approximately equally. Thus, it is easy to distinguish real line emission from
bad weather noise in a spectrum because the real emission is confined to
a narrow frequency range, typically a few channels wide.
However, the traditional integrated intensity display,
shown in Figure \ref{fig1}, appears to be dominated by noisy data: any real
emission is confused with the noisy data.
In order to minimise the visual impact of the striping, we employed a method
that highlighted real emission over the artifacts introduced by bad weather.
For each position-element in the data cube (effectively each pixel in the
integrated intensity map) we calculated the rms noise level and the peak
intensity level of the spectrum at that postion. We then formed a new map
where each element is the ratio of the peak intensity level to the rms noise
level. This effectively creates a map where the brightest pixels indicate
real, narrow-lined emission and can therefore be used as an approximate guide
to the morphology and extent of real emission in the data cube. Emission
maps for G305.5+0.5 and G333.2-0.7 are shown in Figures \ref{fig2a} and
\ref{fig2b}, respectively. Note that because the brightness shown in these maps
depends on a changing rms noise level, the intensity cannot be used to
reliably represent real emission intensity.

We remind the reader that this pilot survey is not designed to produce the
best map of emission possible: that is left for the main HOPS. It is designed
to determine what we can expect from the main survey. HOPS observations will
not be made in such poor weather conditions.

\subsection{G305.5+0.5}
\begin{figure}[h]
\begin{center}
\includegraphics[width=0.48\textwidth, angle=0]{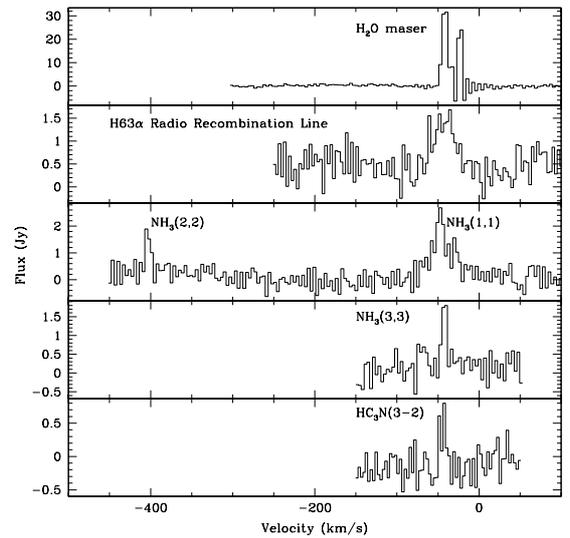}
\caption{Spectra of detected emission lines at G305.20+0.20. From the top
spectrum, running down the page: \water~maser, H63$\alpha$, \ammonia~(1,1)
and (2,2), \ammonia~(3,3) and HC$_3$N(3--2). Note that all radio recombination
lines from H69$\alpha$ to H62$\alpha$ were detected, but only H63$\alpha$ is
shown here.}
\label{fig3}
\end{center}
\end{figure}

The \water~maser emission map in Figure \ref{fig2a} shows
four clear maser detections as black spots in a complex emitting region centred
on G305.2+0.25. The plus symbols in the map
show all \water~masers that were identified by careful inspection of
the full data cube: a peak intensity map was constructed using the peak pixel
in each spectrum at every position in the map. Each pixel 5$\sigma$ above the
rms noise level in this map was visually checked for real maser emission, as
opposed to a noise spike. Real maser emission is distinguished by ocurring
in more than one spectral channel and more than one spatial pixel.
In this field we detect eleven \water~masers, seven of which are new.
The other emission maps shown in Figure
\ref{fig2a} also show emission confined to the bottom right corner.
Figure \ref{fig3} shows spectra of emission
in all detected lines from G305.2+0.2, which is circled in the \water~maser
map in Figure \ref{fig2a}. A summary of detected emission towards each
\water~maser is given in Table \ref{dets}

\begin{table}
\caption{\protect\footnotesize{Detections of other species at the positions
of \water~masers, and previous \water~maser identifications,
for both G305.5+0.5 and G333.2-0.7.
The first two columns denote the position of the \water~masers
in Galactic coordinates. The third column identifies each \water~maser as
a new or previous detection. New \water~masers in G333.2-0.7
marked with an asterisk were observed but
not detected by \citet{breen07}, which is probably due to intrinsic variability
of the masers. The remaining columns are either noted with a `Y',
indicating that emission was detected in the transition or a `N', indicating
no emission was detected at this position.}}
\scriptsize{
\begin{tabular}{cccccccc}
\hline
l & b & \water & \ammonia & \ammonia & \ammonia & H63$\alpha$ & HC$_3$N\\
($^\circ$) & ($^\circ$) &maser&(1,1)&(2,2)&(3,3)&(RRL)&(3--2)\\
\hline
305.13 & +0.08 &new& Y & Y & N & N & N \\
305.20 & -0.02 &a& Y & Y & N & Y & Y \\
305.21 & +0.21 &b& Y & Y & Y & Y & Y \\
305.28 & +0.00 &new& Y & Y & N & Y & N \\
305.33 & +0.07 &new& N & N & N & Y & N \\
305.36 & +0.15 &c& Y & Y & N & Y & N \\
305.37 & +0.22 &d& Y & Y & Y & Y & N \\
305.55 & -0.02 &new& N & N & N & Y & N \\
305.74 & +0.08 &new& N & N & N & N & N \\
305.85 & +0.07 &new& N & N & N & N & N \\
305.89 & +0.03 &new& Y & Y & N & N & N \\
332.68 & -0.28 &new& N & N & N & N & N \\
332.73 & -0.63 &new$^*$& Y & N & N & Y & N \\
332.83 & -0.55 &e& Y & Y & Y & Y & N \\
332.94 & -0.68 &new& Y & Y & Y & N & N \\
332.98 & -0.44 &new& Y & Y & Y & Y & N \\
333.03 & -0.44 &new$^*$& Y & Y & Y & Y & N \\
333.06 & -0.49 &f& Y & Y & N & Y & Y \\
333.13 & -0.43 &c& Y & Y & Y & Y & Y \\
333.22 & -0.41 &f& Y & Y & Y & Y & Y \\
333.60 & -0.23 &d& N & N & N & Y & N \\
\hline
\end{tabular}
a -- \citet{caswell89}; b -- \citet{kaufmann76}; c -- \citet{caswell74}; d -- \citet{johnston72}; e -- \citet{braz82}; f -- \citet{breen07}; 
}
\label{dets}
\end{table}

\subsection{G333.2-0.7}
Emission maps are shown in Figure \ref{fig2b}. The \water~maser emission map
shows ten detected \water~masers (plus symbols), five of which are new
(see Table \ref{dets}). Note that two of the detected masers appear to
fall outside the mapped region. This is because the outer pixels in Figure
\ref{fig2b} are automatically masked by the data processing routines
when creating this emission map. The \water~masers appearing outside
this region were identified in the original \water~maser data cube.

\begin{figure}[h]
\begin{center}
\includegraphics[width=0.48\textwidth, angle=0]{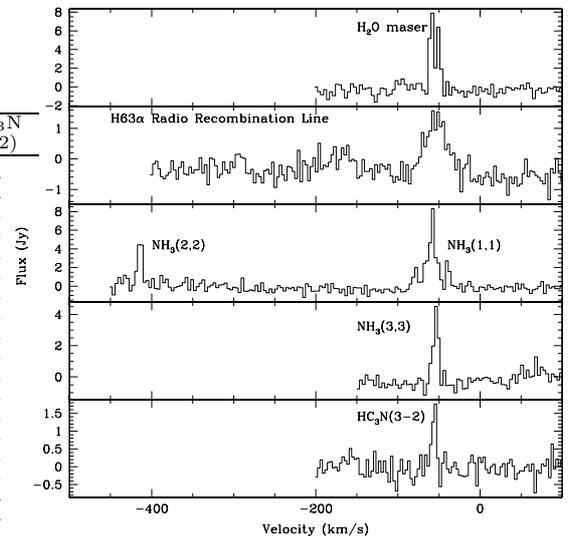}
\caption{Spectra of detected emission lines at G333.14-0.44. From the top
spectrum, running down the page: \water~maser, H63$\alpha$, \ammonia~(1,1)
and (2,2), \ammonia~(3,3) and HC$_3$N(3--2). Note that all radio recombination
lines from H69$\alpha$ to H62$\alpha$ were detected, but only H63$\alpha$ is
shown here.}
\label{fig4}
\end{center}
\end{figure}

\ammonia, HC$_3$N and
H63$\alpha$ appear stronger and more extended than in Figure \ref{fig2a}.
Figure \ref{fig4} shows spectra of detected emission at G333.14-0.44,
represented by the circle in Figure \ref{fig2b}. Again, all lines are
clearly detected. The outer hyperfine components of \ammonia~(1,1) can also
be seen. A summary of detected emission towards each \water~maser is given in
Table \ref{dets}.

\citet{breen07} conducted an untargetted survey of
\water~masers having a substantial overlap with our test region. We note that
\citet{breen07} detected three masers (their sources 6, 7 and 8)
that we did not detect. It is difficult
to identify the reason for this as the broadband mode of HOPS has much wider
channels (3.6\,\kms) than \citet{breen07} (0.25\,\kms), making a direct
comparison of sensitivities dependant on the velocity structure of the emission.
For example, with much broader channels, we are unlikely to detect a single,
weak narrow-lined maser feature, but a maser site comprising
of many weak individual maser peaks, closely spaced in velocity
(within 3.6\,km\,\kms) is more likely to be detected by us.
However, it is likely
that maser variability plays a major part. Variablility is a likely explanation
for the non-detection of source 6 in our pilot survey as this maser did not
appear in followup observations by \citet{breen07}. A more detailed comparison
of previously detected masers and HOPS observations will be conducted during
the main survey, when we use the narrowband mode and can directly compare
narrow lined emission. Due to the inherent variability of \water~masers and
because HOPS will be a single-epoch survey, we will not be able to detect all
\water~masers above the sensitivity limit, but HOPS will give us a snapshot
of maser emission.

\section{Discussion}
\subsection{Weather and Summer Observations}
We intend to make observations for HOPS during the Australian summer, when
demand for telescope time is limited because 3\,mm observing is not possible.
Very few mm-wave observing projects have been carried out outside of the
traditional winter observing season. Thus, it is important that we
understand how the data might be affected by typical summer weather conditions
like high temperatures and cloud cover. During the observations, we encountered
many varied conditions. Our experience indicates that high temperatures do
not seriously affect the data quality when the sky is clear and the humidity
is low. The data are seriously affected when thick cumulus clouds or rain
occurs over the telescope, leading to the stripes seen in Figure \ref{fig1}.

\begin{figure}[h]
\begin{center}
\includegraphics[width=0.48\textwidth, angle=0]{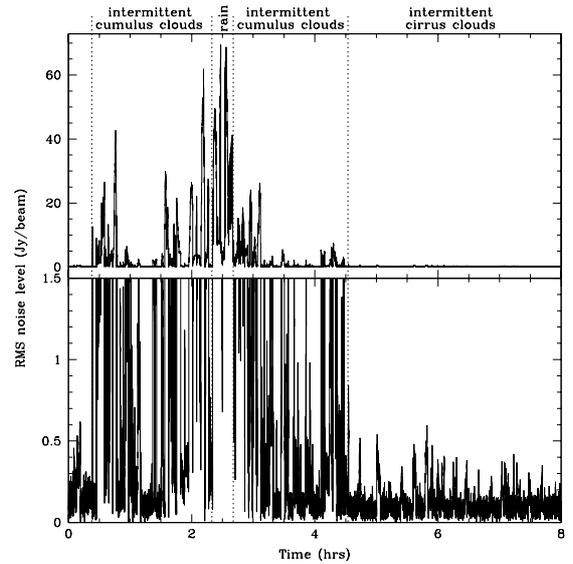}
\caption{RMS noise levels calculated from spectra plotted as a function
of time. The upper plot shows the full rms noise range, whereas the lower plot
shows the same rms noise data scaled to show only the lowest noise data, below
1.5\,Jy/beam. Above the plots are descriptions of the weather at various
time periods separated by the dotted vertical lines.}
\label{fignoise}
\end{center}
\end{figure}

A plot of the rms noise values for a time series of observations is shown in
Figure \ref{fignoise}. During the first 0.5 hours, there was clear weather.
Between approximately 0.5 and 1.0 hours there were cumulus clouds over the
telescope followed by a period of mediocre weather (some clear sky and some
cumulus cloud cover) up to 1.5 hours. After 1.5 hours, cumulus cloud cover
increased until around 2.4 hours when there was rain over the telescope. The
rain continued until about 3.0 hours. The cumulus clouds slowly dissipated
after this until about 4.5 hours, after which,
the weather was mainly clear sky, with
some cirrus cloud over the telescope. The effects of the cirrus clouds can be
seen as minor increases in the rms noise level of a factor of a few
in the bottom plot of Figure \ref{fignoise}.

During the times of cumulus cloud cover or rain, the rms noise level
dramatically increased over the baseline level by factors of up to 400.
The rms noise level was below 0.4\,Jy/beam (three times the baseline level)
only 7\% of the time. We consider these conditions to be bad weather.
Note that the baseline  or nominal good weather
rms noise level in Figure \ref{fignoise} is approximately
0.13\,Jy/beam and can be seen after about 4.5 hours. During the
time after 4.5 hours where there were only cirrus clouds or clear sky over the
telescope the noise is below 0.4\,Jy/beam (three times the
baseline level) for 98\% of the time. We consider these conditions to be good
weather. Such a strong contrast in the rms noise levels between good and bad
weather make it easy to discern between the two.

We found that over the entire
observing run the noise level is below 0.4\,Jy/beam for 70\% of the time.
We also found that there was good weather (ie. cirrus clouds or clear sky)
approximately 70\% of the time, during our time at the telescope, which
included observations for other projects totalling one week.
This figure takes into account observing both during the day and the night,
however we observed that the cumulus clouds tend to form in the afternoons,
with significantly less cumulus clouds during the night and early morning.
We intend to take advantage of this by restricting observations to night and
early morning. Thus, we assume that at least 70\% of the observing time will
be useful for HOPS, when only cirrus or clear sky is over the telescope.
For times when the weather is not good enough for observations, we will flag
out the bad data, reobserve this portion of the sky and substitute this for
the bad data.

We note that the rms noise levels quoted above are based on the \water~maser
data. We found that in comparison, the rms noise levels do not change by more
than 15\% over these levels at different frequencies in the band.

\subsection{\water~Maser Detection Rate}
Over the two fields covered, we detected a total of twenty-one masers,
with eleven in
G305.5+0.5 and ten in G333.2-0.7. As detailed below, we expect to have better
sensitivity to \water~masers in the main HOPS. We therefore expect to detect
more than 1000 masers over ninety square degrees of the Galactic plane,
assuming at least eleven masers per square degree. Based on the number of new
detections in the pilot survey (twelve), we may expect about half of the maser
detections to be new, although we caution that this is based on a very small
sample of detections.

\subsection{Galactic Latitude Distribution}
In order to specify how wide in Galactic latitude HOPS should be, we 
investigated how far off the Galactic plane we can find emission. As can
be seen from Figures \ref{fig2a} and \ref{fig2b}, all the emission (both
thermal and maser) is confined tightly to the Galactic plane and does not extend
further than these maps. For G305.5+0.5, the emission is well confined within
0.5$^\circ$ of the Galactic plane. This is also true for nearly
all of the emission seen in G333.2-0.7. Therefore, we define the HOPS width to
be $\pm 0.5^\circ$ in Galactic latitude, which will allow us to observe most
of the detectable thermal emission in our Galaxy using this setup.
For cases like G333.2-0.7, where emission
appears to extend beyond these limits, we will supplement the HOPS with maps
to cover the extended emission.

Although \water~and Class II methanol masers do not trace identical populations,
there is a good deal of overlap between the two (eg. \citealt{beuther02}).
We can estimate the percentage of \water~masers that
might occur outside our survey region
by assuming a similar Galactic latitude distribution as methanol masers.
\citep{walsh97} targetted a large sample of {\em IRAS} sources and determined
a FWHM distribution of 0.5$^\circ$ for methanol masers. This agrees
with the distribution found in an untargetted, but smaller sample, survey
of methanol masers by \citet{pandian07}. Based on this, approximately
70\% of methanol masers should occur within 0.5$^\circ$ of the Galactic plane.
Therefore, we expect that about 30\% of \water~masers will lie outside
the HOPS survey region.

We note that there is one weak source of \ammonia~(1,1) emission detected in
the G305.5+0.5 map far from the Galactic plane, located at G305.53+0.76.
This will be discussed further in \S\ref{extranh3}.

\subsection{Detected Lines and Spectrometer Mode}
The lines that were detected in this pilot survey are \water~masers,
\ammonia~(1,1), (2,2) and (3,3), radio recombination lines H62$\alpha$
to H69$\alpha$ and HC$_3$N (3--2). Since the multiple
radio recombination lines are largely redundant, we consider that we have
detected six unique lines. As previously mentioned, these pilot
observations were conducted using the broadband mode of MOPS, which covers
the entire 8.4\,GHz band, but yields only moderate velocity resolution of
a few \kms~per channel. We chose to observe in this mode because we did
not know how many lines we might detect in this frequency range.

The MOPS has an alternate mode, called `zoom' mode, in which up to sixteen
137.5\,MHz-wide spectra can be taken simultaneously. These spectra are
divided up into four sets of four spectra, with one set for each 2.2\,GHz
IF. Within each IF, there is virtually unlimited freedom to tune each of
the four spectra. As noted above, we only detected six lines in the pilot
survey, which can be adequately accommodated using the zoom mode. With 4096
channels per spectrum, we will have 0.5\,\kms~per channel resolution
in HOPS. This has the added bonus of effectively increasing the
signal to noise ratio for narrow-lined maser detections, which suffer from
bandwidth smearing in the broadband observations. Smaller bandwidths will
also aid us in sky subtraction and baselining.

\begin{figure}[h]
\begin{center}
\includegraphics[width=0.48\textwidth, angle=0]{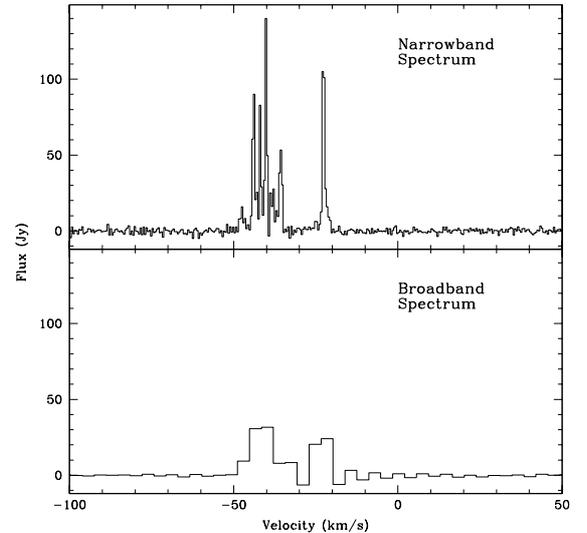}
\caption{\water~maser spectra for {\bf (top)} narrowband zoom mode and
{\bf (bottom)} broadband mode plotted using the same scaling.
The narrowband spectrum shows more detail in the emission
features and allows the detection of weaker masers in the spectrum.}
\label{masercomp}
\end{center}
\end{figure}

We repeated
a small map around G305.20+0.20, whose spectra are shown in Figure \ref{fig3}.
The small map used exactly the same settings (eg. on-the-fly speed, reference
position) as the main G305.5+0.5 map, except that the zoom mode was used.
Comparison spectra of the detected \water~maser are shown in Figure
\ref{masercomp}. It is clear that the zoom mode allows the detection of more
maser emission structure in the spectrum, and that the maser features are
more prominent, making their detection easier.

Using the broadband mode, the weakest maser we detected is G305.13+0.08, which
has a peak flux density of 2.4\,Jy. This is approximately a 5$\sigma$ detection
at this position in the map and gives a guide as to the detection limit
in these observations. Given typical line widths of maser emission (about
1\,\kms~or less), we expect to achieve a factor of 2-3 improvement in
sensitivity with the zoom mode. Since thermal lines tend to be a few \kms,
it is expected that we may get a marginal improvement in sensitivity in
some regions with line widths 2-3\,\kms, but no improvement for line
widths of 3-5\,\kms.

\subsection{G305.53+0.76 - a new \ammonia~cloud}
\label{extranh3}
\begin{figure}[h]
\begin{center}
\includegraphics[width=0.48\textwidth, angle=0]{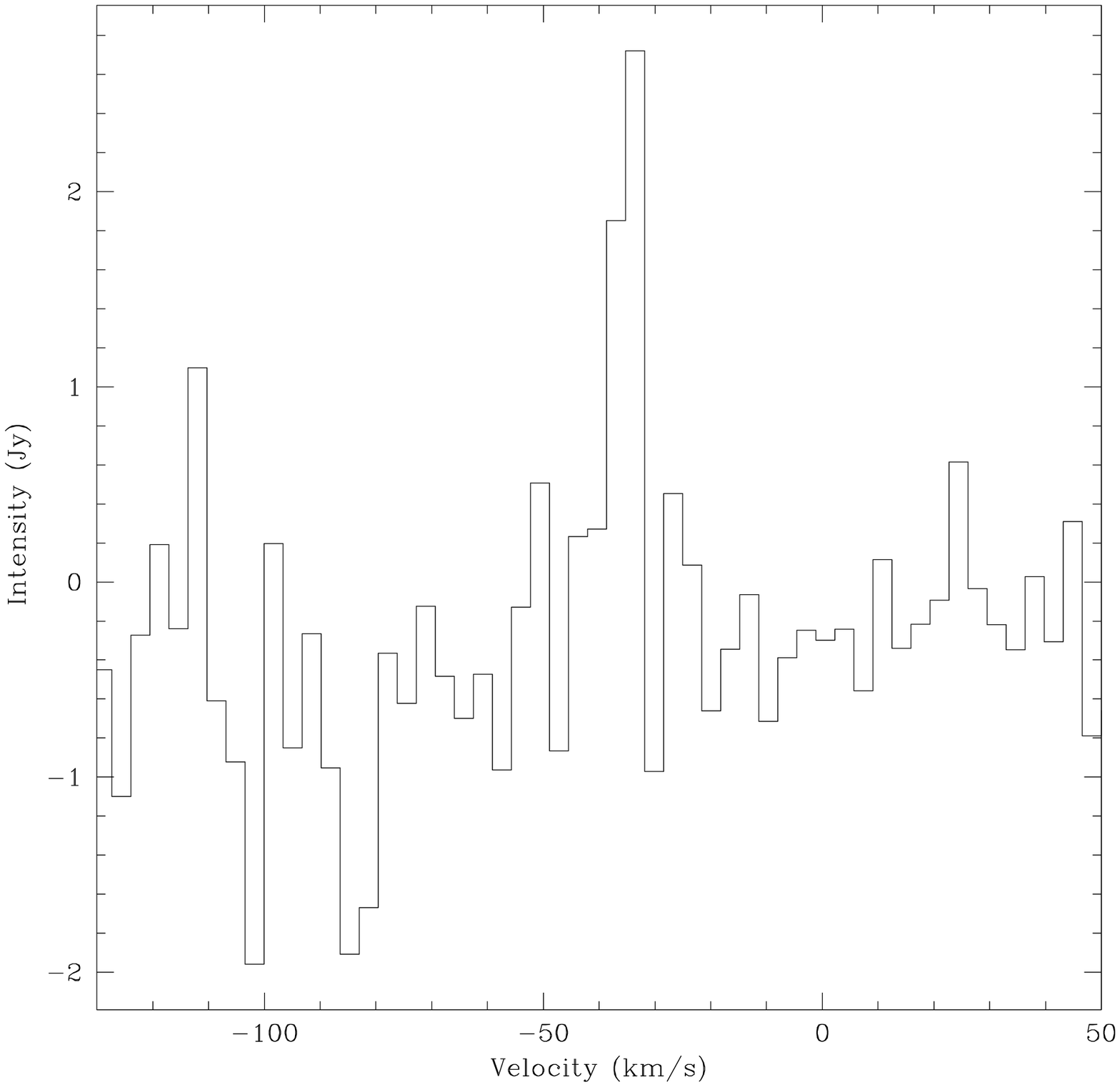}
\includegraphics[width=0.48\textwidth, angle=0]{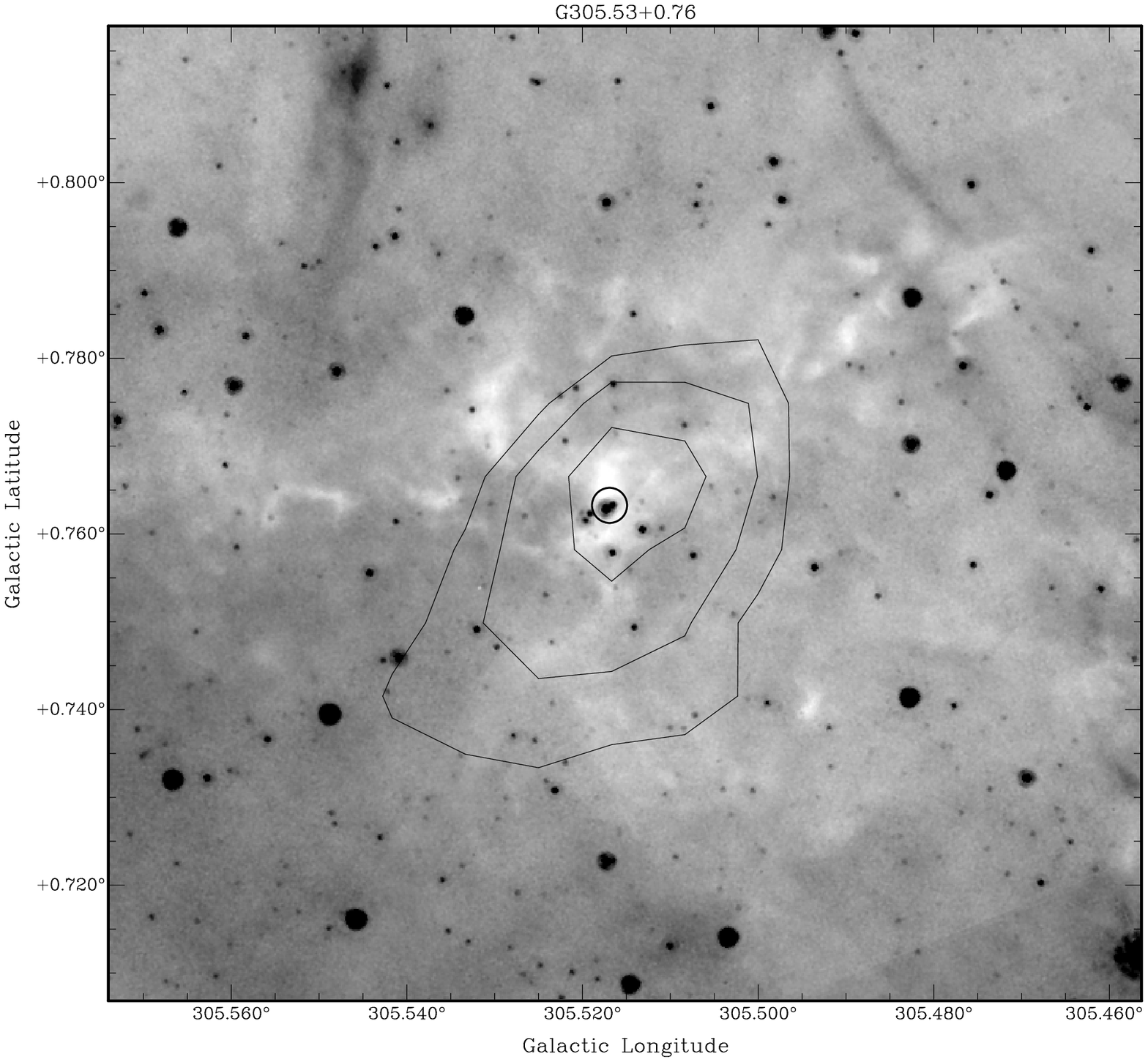}
\caption{{\bf (top)} \ammonia~(1,1) spectrum for G305.53+0.76, showing
emission at -34\,\kms. {\bf (bottom)} Contours show \ammonia~(1,1) emission,
superimposed on the greyscale which
shows 8.0\,$\mu$m emission from the GLIMPSE. An infrared dark cloud is seen
to coincide with the peak of the \ammonia~emission. The circle highlights
an infrared source which shows unusually red colours, implying that it is
deeply embedded.}
\label{30553spec}
\end{center}
\end{figure}

We detected a weak \ammonia~(1,1) feature far removed from the bulk of
\ammonia~emission in the G305.5+0.5 map at G305.53+0.76. The emission is not
obvious in Figure \ref{fig2a} because it lies close to a noisy part of the
data cube. The spectrum of this source is shown in the top half of
Figure \ref{30553spec}. The bottom half shows the contours of \ammonia~emission
overlayed on 8.0\,$\mu$m emission, shown as greyscale. The 8.0\,$\mu$m emission
was taken as part of GLIMPSE (The Galactic Legacy Infrared Mid-Plane Survey
Extraordinaire) with the {\em Spitzer Space Telescope} \citep{benjamin03}.
The GLIMPSE 8.0\,$\mu$m image clearly shows an infrared dark cloud coincident
with the peak of \ammonia~emission. Such a dark cloud suggests that
this is a site of star formation. This is supported by the fact that the
star found in the darkest part of the cloud (circled in Figure \ref{30553spec})
has very red colours, when compared to the other GLIMPSE bands. This is
indicative of a deeply embedded object.

Based on the \ammonia~integrated intensity (33\,K\,\kms), we can estimate the
mass of the cloud. Note that we have assumed a beam efficiency of 0.5 for
Mopra at 12\,mm, but this value is not well characterised, resulting in some
uncertainty in the integrated intensity.
There is a tentative (2.5$\sigma$) detection of \ammonia~(2,2) which we have
used to estimate the kinetic temperature. We determine the kinetic temperature
from the rotational temperature, using the standard formulation for
\ammonia~(1,1) and (2,2) spectra \citep{ho83}. The kinetic temperature has
an upper limit of 22\,K and so we use 22\,K
as an estimate of the temperature in determining the column
density. To estimate the \ammonia~column density, we use equations 1 and 2
of \citet{walsh06}. Here we assume the dipole moment for \ammonia~is
$4.9 \times 10^{30}$C\,m$^{-1}$ (1.48 Debye). The derived \ammonia~column
density is $4.9 \times 10^{19}$\,m$^{-2}$.

\citet{pillai06} find an average
abundance ratio of $\chi_{{\rm NH}_3} = N({\rm NH}_3)/N({\rm H}_2) = 4 \times
10^{-8}$ for infrared dark clouds, and so we adopt this value here. We assume
a distance of 2.5\,kpc to G305.53+0.76 based on the near kinematic distance.
We choose the near kinematic distance because the infrared dark cloud is
seen in absorption against the Galactic background emission.
However, we caution that we do not consider this a robust
distance due to the inherent
uncertainty in kinematic distances. Thus, we only derive an approximate mass
for G305.53+0.76 of 200M$_\odot$; this mass could vary by a factor
of a few, given the uncertainties that go into the assumptions of the various
factors. This mass is much smaller than those found in giant molecular clouds
where high mass stars are thought to form. It is not clear whether or not
this region will form low mass or high mass or any stars,
but is worthy of further investigation to determine its nature.

\section{HOPS survey design}
Having completed the pilot survey, we can now outline the main HOPS. We intend
to survey the regions from l=300$^\circ$ through to l=0$^\circ$
(the Galactic centre) and on to l=30$^\circ$. Galactic latitudes of
b=+0.5$^\circ$ to b=-0.5$^\circ$ will be covered.
This will result in a total of ninety square degrees surveyed. Therefore, we
will concentrate on the busier regions of the inner Galaxy. We will
observe the following lines as our main targets: \water~masers, \ammonia~(1,1),
(2,2) and (3,3), H63$\alpha$ and HC$_3$N (3--2), using the zoom mode,
which will allow us to detect weaker masers than in the pilot survey.
Given these
specifications, and that we have detected twenty-one masers in two square
degrees, we expect to detect at least
1000 \water~masers and a few hundred sources of each of the other tracers in
HOPS. The spare zoom windows will be used to look for emission that may
sporadically occur, such as non-metastable masing transitions of \ammonia,
Class I and Class II CH$_3$OH masers.

It is hoped that HOPS will yield a large sample of \water~masers that can be
directly compared to other masing species (eg. 6.7\,GHz CH$_3$OH Class II
masers) with a statistically unbiased sample. In addition, the \ammonia~emission
can be used to map out the spiral structure of the southern Galaxy and tell us
about the high-density structure of our Galaxy (n$>10^5$\,cm$^{-3}$). Such
a high density survey can then be compared to lower density surveys using
$^{12}$CO (n$\sim10^2$\,cm$^{-3}$) \citep{mizuno01}.


\section*{Acknowledgments} 
We thank the Australian Research Council and UNSW, The
University of Sydney and Monash University for funding to provide the
UNSW MOPS spectrometer used for these observations. We thank the anonymous
referee whose comments have greatly improved the quality of this paper.



\begin{thebibliography}{}
\bibitem[Bains et al.(2006)]{bains06}Bains, I. et al. 2006, MNRAS, 367, 1609
\bibitem[Barlow et al.(1996)]{barlow96}Barlow, M.~J. et al. 1996, A\&AL, 315, 241
\bibitem[Braatz et al.(1996)]{braatz96}Braatz, J.~A., Wilson, A.~S. \& Henkel, C. 1996, ApJS, 106, 51
\bibitem[Braz \& Scalise(1982)]{braz82}Braz, M.~A. \& Scalise, E.~Jr. 1982, A\&A, 107, 272
\bibitem[Benjamin et al.(2003)]{benjamin03}Benjamin R.~A. et al. 2003, PASP, 115, 953
\bibitem[Bergin et al.(2006)]{bergin06}Bergin, E.~A., Maret, S., van der Tak, F.~F.~S., Alves, J., Carmody, S.~M. \& Lada, C.~J. 2006, ApJ, 645, 369
\bibitem[Beuther et al.(2002)]{beuther02}Beuther, H., Walsh, A., Schilke, P., Sridharan, T.~K., Menten, K.~M. \& Wyrowski, F. 2002, A\&A, 390, 289
\bibitem[Breen et al.(2007)]{breen07}Breen, S.~L. et al. 2007, MNRAS, 377, 491
\bibitem[Caswell et al.(1989)]{caswell89}Caswell, J.~L., Batchelor, R.~A., forster, J.~R. \& Wellington, K. J. 1989, AuJPh, 42, 331
\bibitem[Caswell et al.(1974)]{caswell74}Caswell, J.~L., Batchelor, R.~A., Haynes, R.~F. \& Huchtmeier, W.~K. 1974, AuJPh, 27, 417
\bibitem[Cohen et al.(2007)]{cohen07}Cohen, R.~J. et al. 2007, in Triggered Star Formation in a Turbulent ISM (IAU237), Eds. B. G. Elmegreen \& J. Palous, (Cambridge: Cambridge University Press), 403
\bibitem[Claussen et al.(1984)]{claussen84}Claussen, M.~J. et al. 1984, ApJL, 285, 79
\bibitem[Genzel \& Downes(1979)]{genzel79}Genzel, R. \& Downes, D. 1979, A\&A, 72, 234
\bibitem[Hinkle \& Barnes(1979)]{hinkle79}Hinkle, K.~H. \& Barnes, T.~G. 1979, ApJ, 227, 923
\bibitem[Ho \& Townes(1983)]{ho83}Ho, P.~T.~P. \& Townes, C.~H. 1983, ARA\&A, 21, 239
\bibitem[Johnston et al.(1972)]{johnston72}Johnston, K.~J., Robinson, B.~J., Caswell, J.~L. \& Batchelor, R.~A. 1972, ApL, 10, 93
\bibitem[Kaufmann et al.(1976)]{kaufmann76}Kaufmann, P. et al. 1976, Nature, 260, 306
\bibitem[Pandian \& Goldsmith(2007)]{pandian07} Pandian, J. \& Goldsmith, P.~F. 2007, ApJ, 669, 435
\bibitem[Pillai et al.(2006)]{pillai06}Pillai, T., Wyrowski, F., Carey, S.~J. \& Menten, K.~M. 2006, A\&A, 450, 569
\bibitem[Miranda et al.(2001)]{miranda01}Miranda, L.~F., G\'{o}mez, Y., Anglada, G. \& Torrelles, J.~M. 2001, Nature, 414, 284
\bibitem[Mizuno et al.(2001)]{mizuno01}Mizuno, A. et al. 2001, PASJ, 53, 1071
\bibitem[Walsh \& Burton(2006)]{walsh06}Walsh, A.~J. \& Burton, M.~G. 2006, MNRAS, 365, 321
\bibitem[Walsh et al.(1997)]{walsh97}Walsh, A.~J., Hyland, A.~R., Robinson, G. \& Burton, M.~G. 1997, MNRAS, 291, 261
\end{thebibliography}
\end{document}